\documentclass[%
 aip,
 jmp,%
 amsmath,amssymb,
 reprint,%
]{revtex4-1}

\usepackage{listings}

\usepackage{subfigure}
\usepackage{graphicx}
\graphicspath{{./images_all/AFM_exchange_stability/}}

\usepackage{dcolumn}
\usepackage{bm}

\begin{document}

\preprint{AIP/123-QED}

\title{Colossal stability of antiferromagnetically exchange coupled nanomagnets}

\author{Kuntal Roy}
\email{kuntal@iiserb.ac.in}
\noaffiliation
\affiliation{Department of Electrical Engineering and Computer Science, Indian Institute of Science Education and Research (IISER) Bhopal, Bhopal, Madhya Pradesh 462066, India}


\begin{abstract}
Bistable nanomagnets store a binary bit of information. Exchange coupled nanomagnets can increase the thermal stability at low dimensions. Here we show that the antiferromagnetically (AFM) coupled nanomagnets can be highly stable at low dimensions than that of the ferromagnetically (FM) coupled nanomagnets. By solving stochastic Landau-Lifshitz-Gilbert equation of magnetization dynamics at room temperature, we analyze the stability of the exchange coupled nanomagnets in the presence of correlated, uncorrelated, and anti-correlated noise. The results show that the correlated noise can make the stability of the AFM coupled nanomagnets very high. Such finding will lead to very high-density non-volatile storage and logic devices in our future information processing systems.
\end{abstract}

\maketitle


The invention and development of charge-based transistor electronics has been a story of great success~\cite{shockley,kilby} and it has been materialized by miniaturization, i.e., cramming more and more transistors on a chip following the so-called Moore's law purely based on observation~\cite{moore65}. However, excessive energy dissipation in the transistors limits the further improvement of the charge-based electronics~\cite{RefWorks:211,RefWorks:148,RefWorks:149,roy14_4}. Electron's spin-based counterpart, so-called spintronics has profound potential~\cite{fert,grunberg} to be the replacement of current transistor-based technology in our future energy-efficient information processing systems~\cite{RefWorks:435,RefWorks:812,RefWorks:774,roy11_news,roy16_spin,roy14_3,RefWorks:2588,RefWorks:2631,RefWorks:2633}. Switching the magnetization direction between the two symmetry-equivalent stable states allows us to implement \emph{non-volatile} storage and process the information~\cite{roy13,roy17,RefWorks:812}. 

There has been enormous progress in the field of spintronics and nanomagnetics in recent years with the advent of new materials and phenomena~\cite{roy_nanotech_2017x}, however, it is a formidable challenge to minimize the dimension of the nanomagnets~\cite{roy14_5} to incur less and less area on a chip. This is due to the reason that the thermal stability of the magnetization state depends on the energy barrier between the two stable states and that is proportional to the volume of the nanomagnet. An energy barrier of about 40 \emph{kT} (\emph{k} is the Boltzmann constant and \emph{T} is the temperature) at room temperature (T = 300 K) is required to maintain the stability of the magnetic bits for 10 years~\cite{RefWorks:2630,RefWorks:2629}.

With an eye to achieve higher and higher areal density of magnetic bits, the magnetic recording has made transition from longitudinal to perpendicular magnetic anisotropy~\cite{RefWorks:1354}, and also there has been demonstration of perpendicular interface anisotropy~\cite{RefWorks:774,RefWorks:903}. However, at small lateral dimensions ($\sim$10 nm), the energy barrier is not sufficient to hold the information bits for long as per standard requirement. The thermal stability at low dimensions remained a primary concern as it goes down to $20\, kT$ at room temperature for nanomagnets of $\sim$10 nm lateral dimension~\cite{RefWorks:903}. There have been studies on whether a coupled composite system of magnetic bits can increase the energy barrier and the thermal stability~\cite{RefWorks:1443,RefWorks:1449,RefWorks:1448,RefWorks:1447,RefWorks:1446,RefWorks:1435,RefWorks:1433,RefWorks:1441,RefWorks:1444,RefWorks:2587,RefWorks:1456,RefWorks:1457,RefWorks:1453,RefWorks:1454,RefWorks:1452, RefWorks:1439,RefWorks:1450,RefWorks:1125, RefWorks:1119,RefWorks:1120,roy15_1}.

Here we analyze the thermal stability of exchange coupled nanomagnets using the stochastic Landau-Lifshitz-Gilbert (LLG) equation of magnetization dynamics~\cite{RefWorks:162,RefWorks:161} in the presence of room temperature thermal fluctuations~\cite{RefWorks:186}. We consider both the antiferromagnetically (AFM) and ferromagnetically (FM) exchange coupled nanomagnets. We analyze the critical magnetization dynamics in the presence of room temperature thermal fluctuations when the noise applied on the two nanomagnets are correlated, uncorrelated, and anti-correlated. We find that the AFM coupled nanomagnets in the presence of \emph{correlated} noise can be very effective in increasing the thermal stability of the nanomagnets. It makes a perfect sense in assuming \emph{correlated} noise experienced by the two nanomagnets due to their proximity while in idle mode (but the magnetizations fluctuate at their respective positions due to thermal agitations). While the FM coupled nanomagnets with very high exchange strength would just act like a composite system, the AFM coupled nanomagnets are more crucial to understand and here the insights have been provided using the LLG equation of magnetization dynamics. 

\begin{figure*}
\centering
\includegraphics[width=\textwidth]{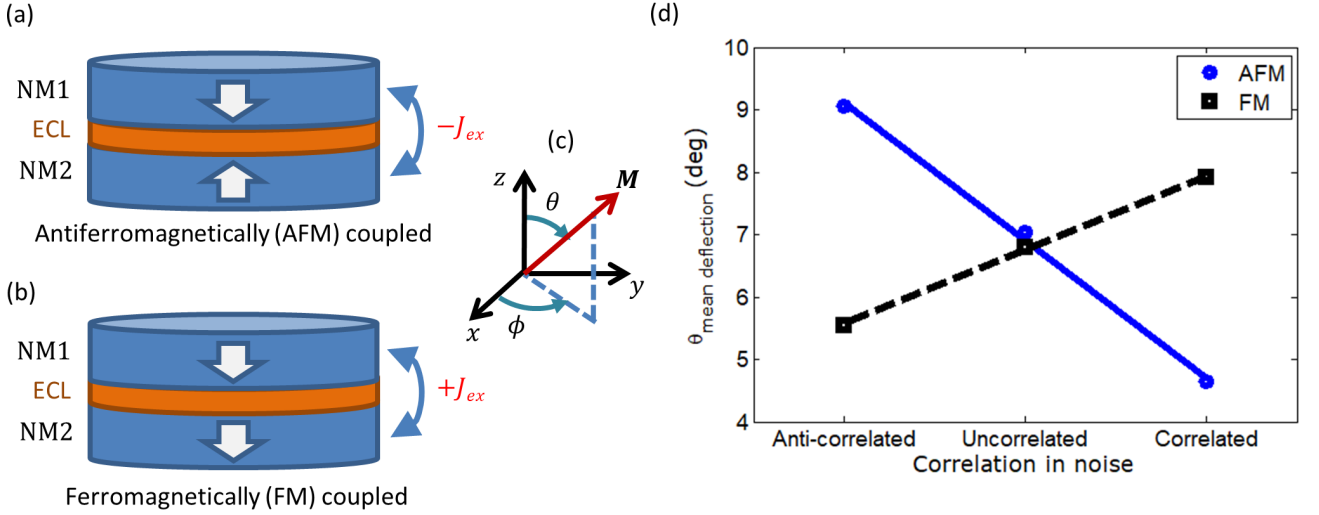}
\caption{\label{fig:noise_correlation} 
Exchange coupled nanomagnets and the mean deflections of magnetizations in the presence of room temperature (300 K) thermal fluctuations, determined from stochastic LLG simulations. (a) Two nanomagnets NM1 and NM2 are \emph{antiferromagnetically} (AFM) exchange coupled via an exchange coupling layer (ECL) with exchange coupling strength $|J_{ex}|$. (b) Two nanomagnets are \emph{ferromagnetically} (FM) coupled via an ECL with exchange coupling strength $J_{ex}$. Here, $J_{ex}$ is positive while $J_{ex}$ is negative for the AFM coupled case. (c) Axes assignment in the standard spherical coordinate system with $\theta$ as polar angle and $\phi$ as azimuthal angle. For the AFM coupled case, NM1 and NM2 are pointing towards $\theta=180^\circ$ and $\theta=0^\circ$, respectively. For the FM-coupled case, both NM1 and NM2 are pointing towards $\theta=180^\circ$. (d) Mean deflections of the magnetizations in the AFM and FM coupled systems in the presence of room temperature thermal noise (correlated, uncorrelated, and anti-correlated) for $|J_{ex}|=1 K_uV$ ($K_uV$ is the uniaxial anisotropy energy barrier). 
}
\end{figure*}

Figures~\ref{fig:noise_correlation}(a) and~(b) show the AFM and FM exchange coupled nanomagnets (NM1 and NM2), respectively separated by an exchange coupling layer (ECL) e.g., Ruthenium~\cite{RefWorks:432,RefWorks:2632} with strength $|J_{ex}|$. The magnetizations $\mathbf{M_{1(2)}}$ of the two coupled nanomagnets have a constant magnitude but a variable direction and thus we can represent it by a vector of unit norm $\mathbf{n_{m,1(2)}} =\mathbf{M_{1(2)}}/|\mathbf{M_{1(2)}}| = \mathbf{\hat{e}_r}$, where $\mathbf{\hat{e}_r}$ is the unit vector in the radial direction in standard spherical coordinate system and the other two unit vectors are $\mathbf{\hat{e}_\theta}$ and $\mathbf{\hat{e}_\phi}$ for $\theta$ and $\phi$ rotations, respectively [Figure~\ref{fig:noise_correlation}(c)]. The potential energy $E_{pot}$ of the nanomagnets can be expressed as the sum of the uniaxial anisotropy energy ($E_{uni}$) and the exchange energy ($E_{ex}$) as
\begin{align}
& E_{pot,1(2)} (\theta_1,\phi_1,\theta_2,\phi_2) \nonumber \\
 & \, = E_{uni,1(2)}(\theta_{1(2)}) + E_{ex}(\theta_1,\phi_1,\theta_2,\phi_2),
\label{eq:energy_total}
\end{align}
where
\begin{equation}
E_{uni,1(2)}(\theta_{1(2)}) = K_{u,1(2)} V_{1(2)} sin^2 \theta_{1(2)},
\label{eq:energy_uni}
\end{equation}
\begin{equation}
E_{ex}(\theta_1,\phi_1,\theta_2,\phi_2) = J_{ex} \mathbf{M_1} \cdot \mathbf{M_2},
\label{eq:energy_ex}
\end{equation}
$K_{u,1(2)}= (1/2) M_{s,1(2)} H_{k,1(2)}$, $K_{u,1(2)}$, $M_{s,1(2)}$, $H_{k,1(2)}$, and $V_{1(2)}$ are the uniaxial anisotropy, the saturation magnetization, the anisotropy field for switching of magnetization, and the volume of the nanomagnets. In the above and onwards, the subscripts $1$ and $2$ denote the cases for the nanomagnets NM1 and NM2, respectively.

The random thermal fluctuations experienced by the nanomagnets are accounted by a random magnetic field 
\begin{equation}
\mathbf{h}(t)= h_x(t)\mathbf{\hat{e}_x} + h_y(t)\mathbf{\hat{e}_y} + h_z(t)\mathbf{\hat{e}_z},
\end{equation}
where $h_i(t)$ ($i \in x,y,z$) are the three components of the random thermal field in Cartesian coordinates. We assume the properties of the random field $\mathbf{h}(t)$ as described in Ref.~\cite{RefWorks:186}. The random thermal field can be written as~\cite{RefWorks:186}
\begin{equation}
h_i(t) = \sqrt{\frac{2 \alpha kT}{|\gamma| M \Delta t}} \; G_{(0,1)}(t) \quad (i \in x,y,z)
\label{eq:ht}
\end{equation}
\noindent
where $\gamma$ is the gyromagnetic ratio for electrons, $\alpha$ is the dimensionless phenomenological Gilbert damping parameter~\cite{RefWorks:162,RefWorks:161}, $M=\mu_0 M_s V$, $1/\Delta t$ is proportional to the attempt frequency of the thermal field, and $G_{(0,1)}(t)$ is a Gaussian distribution with zero mean and unit variance. 

We denote \emph{correlated} and \emph{anti-correlated} noises when $h_{i,1}(t)=h_{i,2}(t)$ and $h_{i,1}(t)=-h_{i,2}(t)$, respectively for $i \in x,y,z$. When there is no correlation between the noises experienced by the two nanomagnets, we term it \emph{uncorrelated}.

\begin{figure*}
\centering
\includegraphics[width=\textwidth]{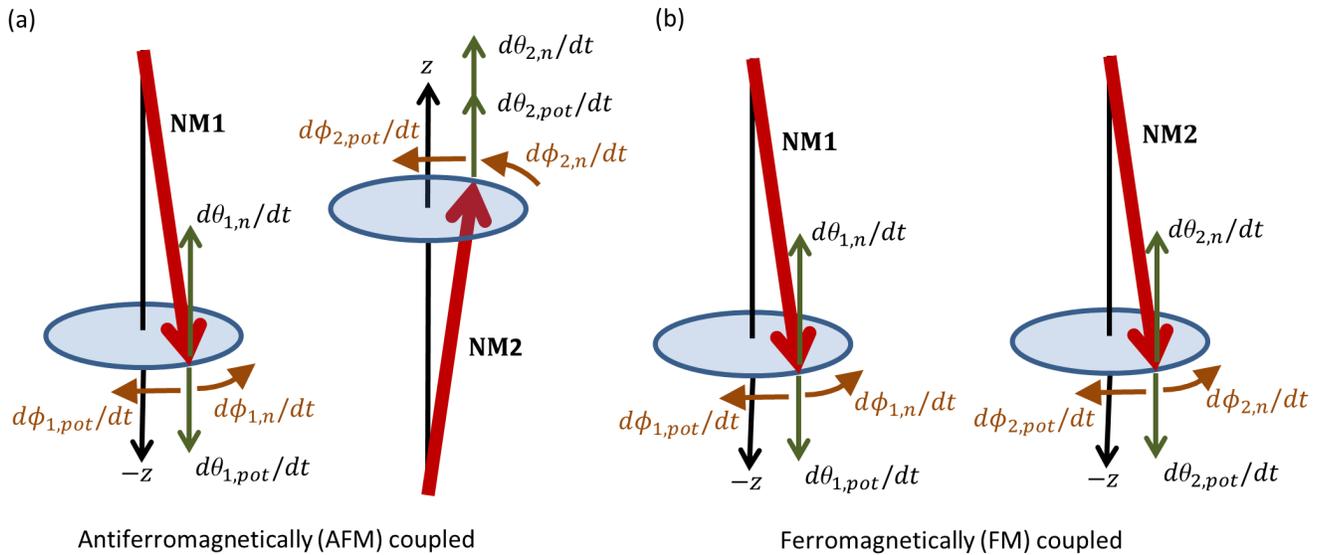}
\caption{\label{fig:correlated_noise_explanation} 
Different motions associated with the AFM and FM coupled nanomagnets. (a) For the AFM coupled case, correlated noise rotates the magnetizations in the same direction (see ${d\phi_{1,n}}/{dt}$ and ${d\phi_{2,n}}/{dt}$, ${d\theta_{1,n}}/{dt}$ and ${d\theta_{2,n}}/{dt}$), which deflects NM1 away from its stable orientation and attracts NM1 towards its stable position. While ${d\phi_{1,n}}/{dt}$ (${d\theta_{1,n}}/{dt}$) counters the motion due to potential energy ${d\phi_{1,pot}}/{dt}$ (${d\theta_{1,pot}}/{dt}$) for NM1, ${d\phi_{2,n}}/{dt}$ (${d\theta_{2,n}}/{dt}$) aids ${d\phi_{2,pot}}/{dt}$ (${d\theta_{2,pot}}/{dt}$) for NM2. (b) For the FM coupled case, correlated noise rotates the magnetizations in the same direction (see ${d\phi_{1,n}}/{dt}$ and ${d\phi_{2,n}}/{dt}$, ${d\theta_{1,n}}/{dt}$ and ${d\theta_{2,n}}/{dt}$), which deflects both the magnetizations NM1 and NM2 away from their stable orientations. Therefore, both ${d\phi_{1,n}}/{dt}$ (${d\theta_{1,n}}/{dt}$) and ${d\phi_{2,n}}/{dt}$ (${d\theta_{2,n}}/{dt}$) counter the motion due to potential energy ${d\phi_{1,pot}}/{dt}$ (${d\theta_{1,pot}}/{dt}$) and ${d\phi_{2,pot}}/{dt}$ (${d\theta_{2,pot}}/{dt}$), respectively. 
}
\end{figure*}

The magnetization dynamics under the action of the torques due to potential energy and thermal noise ($\mathbf{T_{pot}}(t)$ and  $\mathbf{T_{n}}(t)$, respectively) is described by the stochastic Landau-Lifshitz-Gilbert (LLG) equation~\cite{RefWorks:162,RefWorks:161,RefWorks:186} as 
\begin{equation}
\frac{d\mathbf{n_m}}{dt} - \alpha \left(\mathbf{n_m} \times \frac{d\mathbf{n_m}}{dt} \right) = -\frac{|\gamma|}{M} \left\lbrack \mathbf{T_{pot}} +  \mathbf{T_n}\right\rbrack.
\label{eq:LLG}
\end{equation}
We can numerically solve the above equation (using subscripts 1 and 2 for NM1 and NM2, respectively) for $d\theta_{1(2)}/dt$ and $d\phi_{1(2)}/dt$. The dynamics of the rotational angles ($\theta_{1(2)}$,$\phi_{1(2)}$) of NM1 (NM2) are  
\begin{align}
\frac{d\theta_{1(2)}}{dt} &= \left(\frac{d\theta_{{1(2)},uni}}{dt} + \frac{d\theta_{{1(2)},ex}}{dt}\right) + \frac{d\theta_{{1(2)},n}}{dt} \nonumber\\
										&= \frac{d\theta_{{1(2)},pot}}{dt} + \frac{d\theta_{{1(2)},n}}{dt},
\label{eq:dtheta1_dt}
\end{align}
\begin{align}
\frac{d\phi_{1(2)}}{dt} &= \left(\frac{d\phi_{{1(2)},uni}}{dt} + \frac{d\phi_{{1(2)},ex}}{dt}\right) + \frac{d\phi_{{1(2)},n}}{dt} \nonumber\\
&= \frac{d\phi_{{1(2)},pot}}{dt} + \frac{d\phi_{{1(2)},n}}{dt},
\label{eq:dphi1_dt}
\end{align}
where the three components in the two expressions above denote the rotations due to energy barrier caused by uniaxial anisotropy, exchange energy, and the thermal noise, respectively. The first two terms in the expressions can be summed up to denote the rotations $d\theta_{1(2),pot}/dt$ and $d\phi_{1(2),pot}/dt$ due to potential energy of the nanomagnets.

Figure~\ref{fig:noise_correlation}(d) shows the mean deflections of magnetizations for the AFM and FM coupled nanomagnets when $|J_{ex}|=1 K_uV$. The magnetizations are not positioned exactly along the $+z$ axis or $-z$ axis, rather they fluctuate due to thermal agitations. Stochastic LLG equation of magnetization dynamics in the presence of room temperature (300 K) thermal noise is solved to generate these results. Three cases are considered for the characteristics of the thermal noise while applying to the two exchange coupled nanomagnets: (1) Correlated, (2) Uncorrelated, and (3) Anti-correlated. For the AFM coupled system, with the \emph{correlated} noise, the mean deflections of the magnetizations is \emph{less} (i.e., the \emph{stability} is \emph{high}) than that of the other two cases of noise. For the FM coupled nanomagnets, it shows just the opposite trend. For \emph{uncorrelated} noise, AFM and FM coupled systems act similarly, as depicted by the corresponding mean deflections of the magnetizations.

Figure~\ref{fig:correlated_noise_explanation} depicts the explanation behind having \emph{lower} mean deflections of the magnetizations (i.e., \emph{higher} stability) with \emph{correlated} noise for the case of AFM exchange coupled nanomagnets compared to the FM coupled case. The dynamics of the rotations ${d\theta_{1,pot}}/{dt}$, ${d\theta_{1,n}}/{dt}$, ${d\phi_{1,pot}}/{dt}$, ${d\phi_{1,n}}/{dt}$ for NM1 and ${d\theta_{2,pot}}/{dt}$, ${d\theta_{2,n}}/{dt}$, ${d\phi_{2,pot}}/{dt}$, ${d\phi_{2,n}}/{dt}$ for NM2 with \emph{correlated} noise are shown in the Fig.~\ref{fig:correlated_noise_explanation}.

\begin{figure*}
\centering
\includegraphics[width=\textwidth]{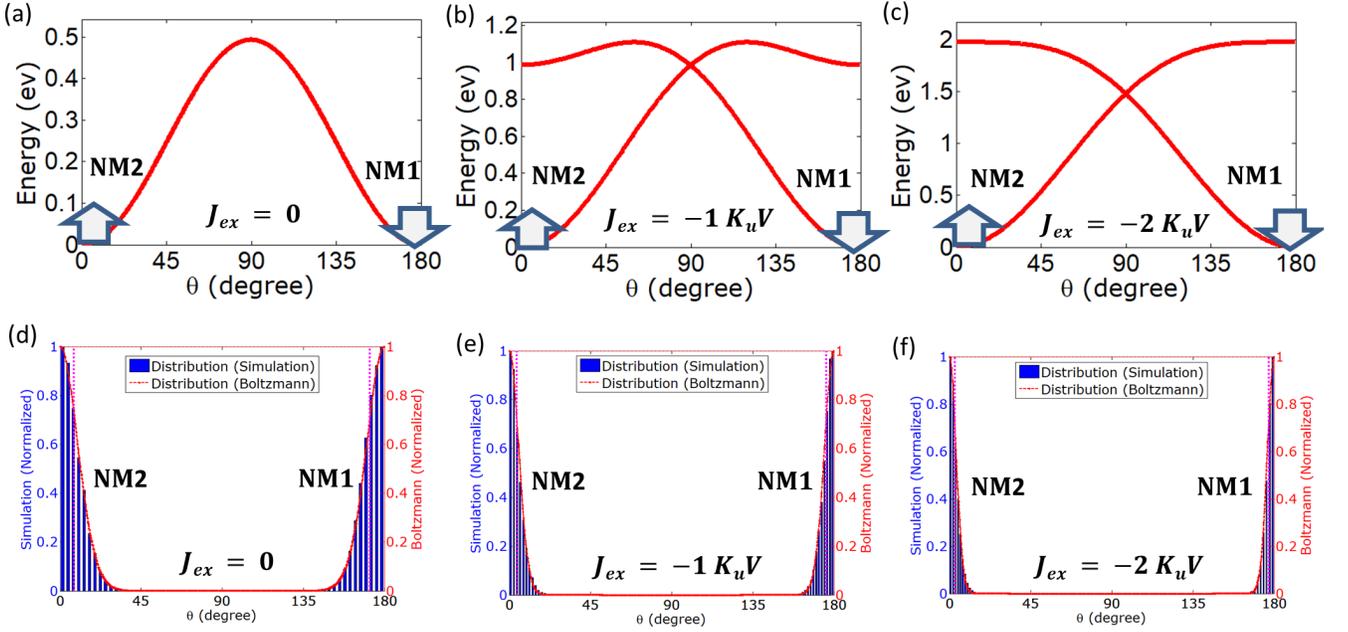}
\caption{\label{fig:distributions_Jex} 
Potential energy landscapes and the stochastic LLG simulation results for the two AFM exchange coupled nanomagnets with \emph{correlated} noise. (a,b,c) Potential energy landscapes when $|J_{ex}|=0, 1\, K_uV,$ and $2\, K_uV$, where $K_uV$ is the energy barrier of the individual nanomagnets. The magnetizations are fluctuating at their respective potential minima due to thermal agitations. (d,e,f) The distributions of the two magnetizations and the fitting with Boltzmann distributions of energy barrier $\sim$20 \emph{kT} ($|J_{ex}|=0$), $\sim$60 \emph{kT} ($|J_{ex}|= 1\, K_uV$), and $\sim$120 \emph{kT} ($|J_{ex}|= 2\, K_uV$), respectively at room temperature (300 K). The mean deflections of the magnetizations are $7.8^o$ ($|J_{ex}|=0$), $4.3^o$ ($|J_{ex}|= 1\, K_uV$), and $3^o$ ($|J_{ex}|= 2\, K_uV$), respectively.}
\end{figure*}

\emph{Correlated noise applied to AFM coupled NM1 and NM2.} Figure~\ref{fig:correlated_noise_explanation}(a) depicts the rotational motions of the  magnetizations NM1 and NM2. NM1 points downward (i.e., $\theta_1 = 180^\circ$), while NM2 points upward (i.e., $\theta_2 = 0^\circ$). Therefore, when the magnetizations are deflected from their positions, the rotational motions $d\phi_{1,pot}/dt$ and $d\phi_{2,pot}/dt$ are in the opposite directions. The other rotational motions $d\theta_{1,pot}/dt$ and $d\theta_{2,pot}/dt$ are in the opposite directions as well so that the magnetizations are attracted towards their respective energy minima. 

Now, at a particular time step, let us assume that thermal noise is attempting to deflect the magnetization NM1 from its stable position, i.e., $d\phi_{1,n}/dt$ due to noise is in the \emph{opposite} direction to $d\phi_{1,pot}/dt$ ($d\theta_{1,n}/dt$ is in the \emph{opposite} direction to $d\theta_{1,pot}/dt$ as well). If the magnetization NM2 faces a \emph{correlated} noise, NM2 experiences the same torque i.e., $d\phi_{2,n}/dt$ due to noise is in the \emph{same} direction to $d\phi_{2,pot}/dt$ ($d\theta_{2,n}/dt$ is in the \emph{same} direction to $d\theta_{2,pot}/dt$ as well). Therefore, NM2 goes toward its stable position (i.e., $\theta_2 = 0^\circ$) rather than deflecting away. Such \emph{counteraction} makes the the AFM exchange coupled system \emph{more} stable with \emph{correlated} noise than that of the case of \emph{uncorrelated} noise. 

If we assume \emph{anti-correlated} noise, $d\phi_{2,n}/dt$ and $d\theta_{2,n}/dt$ would be in the opposite direction than what is shown in the Fig.~\ref{fig:correlated_noise_explanation}(a) and this will allow both the magnetizations NM1 and NM2 to deflect away from their stable positions. Such trend is depicted in the stochastic LLG simulation results for AFM coupled nanomagnets as presented in the Fig.~\ref{fig:noise_correlation}(d).

\emph{Correlated noise applied to FM coupled NM1 and NM2.} For the FM coupled nanomagnets, as depicted in the Fig.~\ref{fig:correlated_noise_explanation}(b), if thermal noise attempts to deflect the magnetization NM1 away from its stable position, it also deflects magnetization NM2 from the stable position for the case of \emph{correlated} noise (see ${d\phi_{1,n}}/{dt}$ and ${d\phi_{2,n}}/{dt}$, ${d\theta_{1,n}}/{dt}$ and ${d\theta_{2,n}}/{dt}$). If the noise is strong enough at some time step, magnetizations can deflect away from their stable orientation ($-z$ axis). Hence, \emph{no counteraction} happens as for the AFM coupled case. Therefore, for the FM coupled nanomagnets, the mean deflections of the magnetizations is \emph{higher} for \emph{correlated} noise than that of case of the \emph{uncorrelated} noise. 

Note that the \emph{anti-correlated} noise for the FM coupled nanomagnets conceptually acts in the very same way as the \emph{correlated} noise acts on the AFM coupled nanomagnets. Such trend is depicted in the stochastic LLG simulation results for FM coupled nanomagnets as presented in the Fig.~\ref{fig:noise_correlation}(d).

\begin{figure*}
\centering
\includegraphics[width=\textwidth]{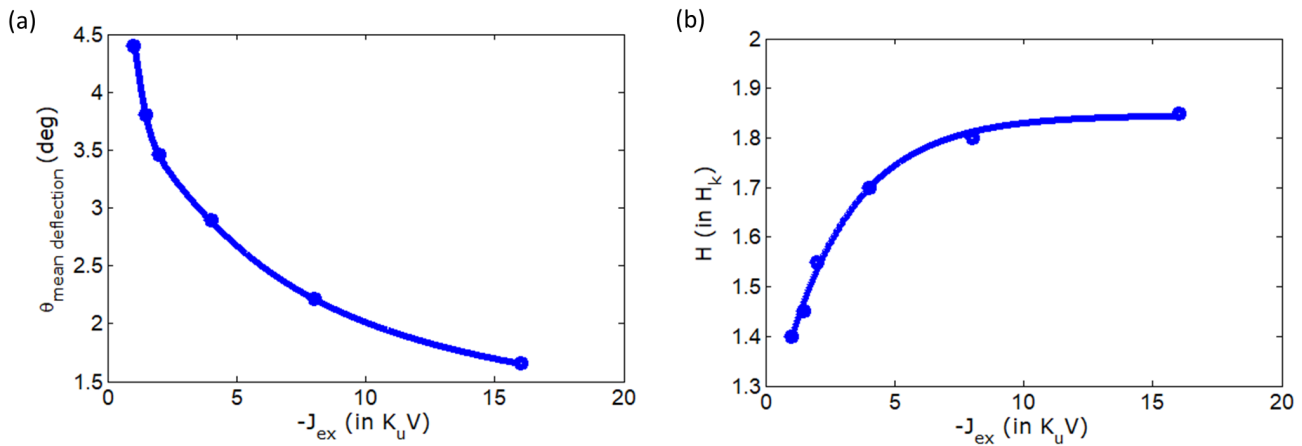}
\caption{\label{fig:stability_H_field} 
Comparison and understanding of the thermal stability with \emph{random} noise field and \emph{deterministic} magnetic field. (a) Mean deflections of the AFM exchange coupled magnetizations with the exchange coupling strength $-J_{ex}$ when \emph{correlated} noise acting on the nanomagnets. (b) The magnetic field $H$ required to switch the AFM exchange coupled nanomagnets with the exchange coupling strength $-J_{ex}$. The field $H$ is applied to one nanomagnet only.}
\end{figure*}

Figure~\ref{fig:distributions_Jex} shows the potential landscapes of the AFM exchange coupled nanomagnets and the distributions of magnetizations due to thermal fluctuations at their potential energy minima. Without any coupling [Figures~\ref{fig:distributions_Jex}(a) and~(d)], the nanomagnets act like isolated ones. Figure~\ref{fig:distributions_Jex}(d) shows the fitting with Boltzmann distribution of $\sim$20 \emph{kT} at room temperature (300 K). This confirms the validity of the simulation results since the nanomagnets chosen have the energy barrier of  $\sim$20 \emph{kT} at 300 K ($M_s=1\, T$, $H_k=2.5\, T$, $\alpha=0.01$, 10 nm diameter with 1 nm thickness). As the exchange coupling strength is increased, the magnetizations become more confined at their potential energy minima [Figures~\ref{fig:distributions_Jex}(b) and~(c)]. With exchange coupling strengths $|J_{ex}|= 1\, K_uV$ and $|J_{ex}|= 2\, K_uV$, the distributions of both the nanomagnets correspond to Boltzmann distributions of $\sim$60 \emph{kT} and $\sim$120 \emph{kT} at room temperature (300 K), respectively. The energy barriers in the fitted Boltzmann distributions can be derived from the mean deflections of the magnetizations $\theta_{mean}$ as $E_b=(1/3)\,(1/sin^2\theta_{mean})$ in the units of \emph{kT}. The exchange coupling strength $|J_{ex}|= 1\, K_uV$ corresponds to $\sim$1 erg/cm$^2$, which can be achieved experimentally~\cite{RefWorks:432,RefWorks:1125}. Therefore, the results clearly demonstrate that the AFM coupled nanomagnets can be highly stable with \emph{correlated} noise. 

Figure~\ref{fig:stability_H_field}(a) depicts that as the exchange coupling strength $|J_{ex}|$ increases, the mean deflection of the magnetizations goes \emph{lower}, i.e., the stability becomes \emph{higher}. As explained earlier, the \emph{inherent} magnetization dynamics of the AFM coupled nanomagnet counters the motion of the magnetizations with \emph{correlated} noise and such counteraction (i.e., stability) increases with $|J_{ex}|$.

Figure~\ref{fig:stability_H_field}(b) shows that as the exchange coupling strength $|J_{ex}|$ increases, the magnetic field $H$ required to switch the AFM exchange coupled nanomagnets does also increase, which is obvious. The required field tends to saturate towards $2\,H_k$, where $H_k$ is the individual switching field for the nanomagnets. Note that a \emph{deterministic} $H$ field is applied at each time step contrary to the case in the Figure~\ref{fig:stability_H_field}(a), where a \emph{random} noise field with mean zero is applied on the nanomagnets. This makes the difference between the switching using a $H$ field and the field due to random thermal noise. It is the noise field (not the $H$ field) that attempts to deflect magnetizations from their stable orientations in real time and the \emph{inherent} dynamics of the AFM exchange coupled nanomagnets determines the stability of the system with \emph{correlated} noise.

To summarize, we have shown by solving stochastic LLG of magnetization dynamics that AFM exchange coupled nanomagnets with \emph{correlated} noise can lead to very high thermal stability at low-dimensions, i.e., the magnetic bits can be scaled down further to achieve higher areal density. With the experimentally feasible exchange strengths, we have shown that the magnetic bits can be highly stable as per the standard requirement of storage. Such finding would allay our concern on the thermal stability of magnetic bits at low-dimensions and would pave the pathway towards ultra high-density non-volatile information storage and logic systems. 

\vspace*{1mm}
The data that support the findings of this study are available from the corresponding author upon reasonable request.


\begin{thebibliography}{52}%
\makeatletter
\providecommand \@ifxundefined [1]{%
 \@ifx{#1\undefined}
}%
\providecommand \@ifnum [1]{%
 \ifnum #1\expandafter \@firstoftwo
 \else \expandafter \@secondoftwo
 \fi
}%
\providecommand \@ifx [1]{%
 \ifx #1\expandafter \@firstoftwo
 \else \expandafter \@secondoftwo
 \fi
}%
\providecommand \natexlab [1]{#1}%
\providecommand \enquote  [1]{``#1''}%
\providecommand \bibnamefont  [1]{#1}%
\providecommand \bibfnamefont [1]{#1}%
\providecommand \citenamefont [1]{#1}%
\providecommand \href@noop [0]{\@secondoftwo}%
\providecommand \href [0]{\begingroup \@sanitize@url \@href}%
\providecommand \@href[1]{\@@startlink{#1}\@@href}%
\providecommand \@@href[1]{\endgroup#1\@@endlink}%
\providecommand \@sanitize@url [0]{\catcode `\\12\catcode `\$12\catcode
  `\&12\catcode `\#12\catcode `\^12\catcode `\_12\catcode `\%12\relax}%
\providecommand \@@startlink[1]{}%
\providecommand \@@endlink[0]{}%
\providecommand \url  [0]{\begingroup\@sanitize@url \@url }%
\providecommand \@url [1]{\endgroup\@href {#1}{\urlprefix }}%
\providecommand \urlprefix  [0]{URL }%
\providecommand \Eprint [0]{\href }%
\providecommand \doibase [0]{http://dx.doi.org/}%
\providecommand \selectlanguage [0]{\@gobble}%
\providecommand \bibinfo  [0]{\@secondoftwo}%
\providecommand \bibfield  [0]{\@secondoftwo}%
\providecommand \translation [1]{[#1]}%
\providecommand \BibitemOpen [0]{}%
\providecommand \bibitemStop [0]{}%
\providecommand \bibitemNoStop [0]{.\EOS\space}%
\providecommand \EOS [0]{\spacefactor3000\relax}%
\providecommand \BibitemShut  [1]{\csname bibitem#1\endcsname}%
\let\auto@bib@innerbib\@empty
\bibitem [{\citenamefont {Shockley}, \citenamefont {Bardeen},\ and\
  \citenamefont {Brattain}(1956)}]{shockley}%
  \BibitemOpen
  \bibfield  {author} {\bibinfo {author} {\bibfnamefont {W.~B.}\ \bibnamefont
  {Shockley}}, \bibinfo {author} {\bibfnamefont {J.}~\bibnamefont {Bardeen}}, \
  and\ \bibinfo {author} {\bibfnamefont {W.~H.}\ \bibnamefont {Brattain}},\
  }\href@noop {} {\emph {\bibinfo {title} {Nobel Lecture in Physics}}}\
  (\bibinfo  {publisher} {The Nobel Foundation, Sweden},\ \bibinfo {year}
  {1956})\BibitemShut {NoStop}%
\bibitem [{\citenamefont {Kilby}(2000)}]{kilby}%
  \BibitemOpen
  \bibfield  {author} {\bibinfo {author} {\bibfnamefont {J.~S.}\ \bibnamefont
  {Kilby}},\ }\href@noop {} {\emph {\bibinfo {title} {Nobel Lecture in
  Physics}}}\ (\bibinfo  {publisher} {The Nobel Foundation, Sweden},\ \bibinfo
  {year} {2000})\BibitemShut {NoStop}%
\bibitem [{\citenamefont {Moore}(1998)}]{moore65}%
  \BibitemOpen
  \bibfield  {author} {\bibinfo {author} {\bibfnamefont {G.~E.}\ \bibnamefont
  {Moore}},\ }\href@noop {} {\bibfield  {journal} {\bibinfo  {journal} {Proc.
  IEEE}\ }\textbf {\bibinfo {volume} {86}},\ \bibinfo {pages} {82} (\bibinfo
  {year} {1998})}\BibitemShut {NoStop}%
\bibitem [{\citenamefont {Borkar}(1999)}]{RefWorks:211}%
  \BibitemOpen
  \bibfield  {author} {\bibinfo {author} {\bibfnamefont {S.}~\bibnamefont
  {Borkar}},\ }\href@noop {} {\bibfield  {journal} {\bibinfo  {journal} {IEEE
  Micro}\ }\textbf {\bibinfo {volume} {19}},\ \bibinfo {pages} {23} (\bibinfo
  {year} {1999})}\BibitemShut {NoStop}%
\bibitem [{\citenamefont {Landauer}(1961)}]{RefWorks:148}%
  \BibitemOpen
  \bibfield  {author} {\bibinfo {author} {\bibfnamefont {R.}~\bibnamefont
  {Landauer}},\ }\href@noop {} {\bibfield  {journal} {\bibinfo  {journal} {IBM
  J. Res. Dev.}\ }\textbf {\bibinfo {volume} {5}},\ \bibinfo {pages} {183}
  (\bibinfo {year} {1961})}\BibitemShut {NoStop}%
\bibitem [{\citenamefont {Keyes}\ and\ \citenamefont
  {Landauer}(1970)}]{RefWorks:149}%
  \BibitemOpen
  \bibfield  {author} {\bibinfo {author} {\bibfnamefont {R.~W.}\ \bibnamefont
  {Keyes}}\ and\ \bibinfo {author} {\bibfnamefont {R.}~\bibnamefont
  {Landauer}},\ }\href@noop {} {\bibfield  {journal} {\bibinfo  {journal} {IBM
  J. Res. Dev.}\ }\textbf {\bibinfo {volume} {14}},\ \bibinfo {pages} {152}
  (\bibinfo {year} {1970})}\BibitemShut {NoStop}%
\bibitem [{\citenamefont {Roy}(2014{\natexlab{a}})}]{roy14_4}%
  \BibitemOpen
  \bibfield  {author} {\bibinfo {author} {\bibfnamefont {K.}~\bibnamefont
  {Roy}},\ }\href@noop {} {\bibfield  {journal} {\bibinfo  {journal} {J. Phys.:
  Condens. Matter}\ }\textbf {\bibinfo {volume} {26}},\ \bibinfo {pages}
  {492203} (\bibinfo {year} {2014}{\natexlab{a}})}\BibitemShut {NoStop}%
\bibitem [{\citenamefont {Fert}(2007)}]{fert}%
  \BibitemOpen
  \bibfield  {author} {\bibinfo {author} {\bibfnamefont {A.}~\bibnamefont
  {Fert}},\ }\href@noop {} {\emph {\bibinfo {title} {The origin, development
  and future of spintronics}}}\ (\bibinfo  {publisher} {Nobel Lecture in
  Physics, The Nobel Foundation, Sweden},\ \bibinfo {year} {2007})\BibitemShut
  {NoStop}%
\bibitem [{\citenamefont {Gr\"{u}nberg}(2007)}]{grunberg}%
  \BibitemOpen
  \bibfield  {author} {\bibinfo {author} {\bibfnamefont {P.}~\bibnamefont
  {Gr\"{u}nberg}},\ }\href@noop {} {\emph {\bibinfo {title} {{From spinwaves to
  giant magnetoresistance (GMR) and beyond}}}}\ (\bibinfo  {publisher} {Nobel
  Lecture in Physics, The Nobel Foundation, Sweden},\ \bibinfo {year}
  {2007})\BibitemShut {NoStop}%
\bibitem [{\citenamefont {Chappert}, \citenamefont {Fert},\ and\ \citenamefont
  {Dau}(2007)}]{RefWorks:435}%
  \BibitemOpen
  \bibfield  {author} {\bibinfo {author} {\bibfnamefont {C.}~\bibnamefont
  {Chappert}}, \bibinfo {author} {\bibfnamefont {A.}~\bibnamefont {Fert}}, \
  and\ \bibinfo {author} {\bibfnamefont {F.~N.~V.}\ \bibnamefont {Dau}},\
  }\href@noop {} {\bibfield  {journal} {\bibinfo  {journal} {Nature Mater.}\
  }\textbf {\bibinfo {volume} {6}},\ \bibinfo {pages} {813} (\bibinfo {year}
  {2007})}\BibitemShut {NoStop}%
\bibitem [{\citenamefont {Kajiwara}\ \emph {et~al.}(2010)\citenamefont
  {Kajiwara}, \citenamefont {Harii}, \citenamefont {Takahashi}, \citenamefont
  {Ohe}, \citenamefont {Uchida}, \citenamefont {Mizuguchi}, \citenamefont
  {Umezawa}, \citenamefont {Kawai}, \citenamefont {Ando}, \citenamefont
  {Takanashi}, \citenamefont {Maekawa1},\ and\ \citenamefont
  {Saitoh}}]{RefWorks:812}%
  \BibitemOpen
  \bibfield  {author} {\bibinfo {author} {\bibfnamefont {Y.}~\bibnamefont
  {Kajiwara}}, \bibinfo {author} {\bibfnamefont {K.}~\bibnamefont {Harii}},
  \bibinfo {author} {\bibfnamefont {S.}~\bibnamefont {Takahashi}}, \bibinfo
  {author} {\bibfnamefont {J.}~\bibnamefont {Ohe}}, \bibinfo {author}
  {\bibfnamefont {K.}~\bibnamefont {Uchida}}, \bibinfo {author} {\bibfnamefont
  {M.}~\bibnamefont {Mizuguchi}}, \bibinfo {author} {\bibfnamefont
  {H.}~\bibnamefont {Umezawa}}, \bibinfo {author} {\bibfnamefont
  {H.}~\bibnamefont {Kawai}}, \bibinfo {author} {\bibfnamefont
  {K.}~\bibnamefont {Ando}}, \bibinfo {author} {\bibfnamefont {K.}~\bibnamefont
  {Takanashi}}, \bibinfo {author} {\bibfnamefont {S.}~\bibnamefont {Maekawa1}},
  \ and\ \bibinfo {author} {\bibfnamefont {E.}~\bibnamefont {Saitoh}},\
  }\href@noop {} {\bibfield  {journal} {\bibinfo  {journal} {Nature}\ }\textbf
  {\bibinfo {volume} {464}},\ \bibinfo {pages} {262} (\bibinfo {year}
  {2010})}\BibitemShut {NoStop}%
\bibitem [{\citenamefont {Ikeda}\ \emph {et~al.}(2010)\citenamefont {Ikeda},
  \citenamefont {Miura}, \citenamefont {Yamamoto}, \citenamefont {Mizunuma},
  \citenamefont {Gan}, \citenamefont {Endo}, \citenamefont {Kanai},
  \citenamefont {Hayakawa}, \citenamefont {Matsukura},\ and\ \citenamefont
  {Ohno}}]{RefWorks:774}%
  \BibitemOpen
  \bibfield  {author} {\bibinfo {author} {\bibfnamefont {S.}~\bibnamefont
  {Ikeda}}, \bibinfo {author} {\bibfnamefont {K.}~\bibnamefont {Miura}},
  \bibinfo {author} {\bibfnamefont {H.}~\bibnamefont {Yamamoto}}, \bibinfo
  {author} {\bibfnamefont {K.}~\bibnamefont {Mizunuma}}, \bibinfo {author}
  {\bibfnamefont {H.~D.}\ \bibnamefont {Gan}}, \bibinfo {author} {\bibfnamefont
  {M.}~\bibnamefont {Endo}}, \bibinfo {author} {\bibfnamefont {S.}~\bibnamefont
  {Kanai}}, \bibinfo {author} {\bibfnamefont {J.}~\bibnamefont {Hayakawa}},
  \bibinfo {author} {\bibfnamefont {F.}~\bibnamefont {Matsukura}}, \ and\
  \bibinfo {author} {\bibfnamefont {H.}~\bibnamefont {Ohno}},\ }\href@noop {}
  {\bibfield  {journal} {\bibinfo  {journal} {Nature Mater.}\ }\textbf
  {\bibinfo {volume} {9}},\ \bibinfo {pages} {721} (\bibinfo {year}
  {2010})}\BibitemShut {NoStop}%
\bibitem [{\citenamefont {Roy}, \citenamefont {Bandyopadhyay},\ and\
  \citenamefont {Atulasimha}(2011)}]{roy11_news}%
  \BibitemOpen
  \bibfield  {author} {\bibinfo {author} {\bibfnamefont {K.}~\bibnamefont
  {Roy}}, \bibinfo {author} {\bibfnamefont {S.}~\bibnamefont {Bandyopadhyay}},
  \ and\ \bibinfo {author} {\bibfnamefont {J.}~\bibnamefont {Atulasimha}},\
  }\href@noop {} {\bibfield  {journal} {\bibinfo  {journal} {Appl. Phys.
  Lett.}\ }\textbf {\bibinfo {volume} {99}},\ \bibinfo {pages} {063108}
  (\bibinfo {year} {2011})},\ \bibinfo {note} {\\News: ``Switching up spin,''
  \it{Nature} \bf{476}\normalfont, 375 (Aug. 25, 2011),
  doi:10.1038/476375c}\BibitemShut {NoStop}%
\bibitem [{\citenamefont {Roy}(2016)}]{roy16_spin}%
  \BibitemOpen
  \bibfield  {author} {\bibinfo {author} {\bibfnamefont {K.}~\bibnamefont
  {Roy}},\ }\href@noop {} {\bibfield  {journal} {\bibinfo  {journal} {SPIN}\
  }\textbf {\bibinfo {volume} {6}},\ \bibinfo {pages} {1630001} (\bibinfo
  {year} {2016})}\BibitemShut {NoStop}%
\bibitem [{\citenamefont {Roy}(2014{\natexlab{b}})}]{roy14_3}%
  \BibitemOpen
  \bibfield  {author} {\bibinfo {author} {\bibfnamefont {K.}~\bibnamefont
  {Roy}},\ }\href@noop {} {\bibfield  {journal} {\bibinfo  {journal} {J. Phys.
  D: Appl. Phys.}\ }\textbf {\bibinfo {volume} {47}},\ \bibinfo {pages}
  {422001} (\bibinfo {year} {2014}{\natexlab{b}})}\BibitemShut {NoStop}%
\bibitem [{\citenamefont {Ramaswamy}\ \emph {et~al.}(2018)\citenamefont
  {Ramaswamy}, \citenamefont {Lee}, \citenamefont {Cai},\ and\ \citenamefont
  {Yang}}]{RefWorks:2588}%
  \BibitemOpen
  \bibfield  {author} {\bibinfo {author} {\bibfnamefont {R.}~\bibnamefont
  {Ramaswamy}}, \bibinfo {author} {\bibfnamefont {J.~M.}\ \bibnamefont {Lee}},
  \bibinfo {author} {\bibfnamefont {K.}~\bibnamefont {Cai}}, \ and\ \bibinfo
  {author} {\bibfnamefont {H.}~\bibnamefont {Yang}},\ }\href@noop {} {\bibfield
   {journal} {\bibinfo  {journal} {Appl. Phys. Rev.}\ }\textbf {\bibinfo
  {volume} {5}},\ \bibinfo {pages} {031107} (\bibinfo {year}
  {2018})}\BibitemShut {NoStop}%
\bibitem [{\citenamefont {Tokura}, \citenamefont {Yasuda},\ and\ \citenamefont
  {Tsukazaki}(2019)}]{RefWorks:2631}%
  \BibitemOpen
  \bibfield  {author} {\bibinfo {author} {\bibfnamefont {Y.}~\bibnamefont
  {Tokura}}, \bibinfo {author} {\bibfnamefont {K.}~\bibnamefont {Yasuda}}, \
  and\ \bibinfo {author} {\bibfnamefont {A.}~\bibnamefont {Tsukazaki}},\
  }\href@noop {} {\bibfield  {journal} {\bibinfo  {journal} {Nature Rev.
  Phys.}\ }\textbf {\bibinfo {volume} {1}},\ \bibinfo {pages} {126} (\bibinfo
  {year} {2019})}\BibitemShut {NoStop}%
\bibitem [{\citenamefont {Puebla}\ \emph {et~al.}(2020)\citenamefont {Puebla},
  \citenamefont {Kim}, \citenamefont {Kondou},\ and\ \citenamefont
  {Otani}}]{RefWorks:2633}%
  \BibitemOpen
  \bibfield  {author} {\bibinfo {author} {\bibfnamefont {J.}~\bibnamefont
  {Puebla}}, \bibinfo {author} {\bibfnamefont {J.}~\bibnamefont {Kim}},
  \bibinfo {author} {\bibfnamefont {K.}~\bibnamefont {Kondou}}, \ and\ \bibinfo
  {author} {\bibfnamefont {Y.}~\bibnamefont {Otani}},\ }\href@noop {}
  {\bibfield  {journal} {\bibinfo  {journal} {Comm. Mater.}\ }\textbf {\bibinfo
  {volume} {1}},\ \bibinfo {pages} {1} (\bibinfo {year} {2020})}\BibitemShut
  {NoStop}%
\bibitem [{\citenamefont {Roy}(2013)}]{roy13}%
  \BibitemOpen
  \bibfield  {author} {\bibinfo {author} {\bibfnamefont {K.}~\bibnamefont
  {Roy}},\ }\href@noop {} {\bibfield  {journal} {\bibinfo  {journal} {Appl.
  Phys. Lett.}\ }\textbf {\bibinfo {volume} {103}},\ \bibinfo {pages} {173110}
  (\bibinfo {year} {2013})}\BibitemShut {NoStop}%
\bibitem [{\citenamefont {Roy}(2017{\natexlab{a}})}]{roy17}%
  \BibitemOpen
  \bibfield  {author} {\bibinfo {author} {\bibfnamefont {K.}~\bibnamefont
  {Roy}},\ }\href@noop {} {\bibfield  {journal} {\bibinfo  {journal} {IEEE
  Trans. Nanotech.}\ }\textbf {\bibinfo {volume} {16}},\ \bibinfo {pages} {333}
  (\bibinfo {year} {2017}{\natexlab{a}})}\BibitemShut {NoStop}%
\bibitem [{\citenamefont {Roy}(2017{\natexlab{b}})}]{roy_nanotech_2017x}%
  \BibitemOpen
  \bibfield  {author} {\bibinfo {author} {\bibfnamefont {K.}~\bibnamefont
  {Roy}},\ }in\ \href@noop {} {\emph {\bibinfo {booktitle} {TechConnect Briefs
  (NanoTech) 2017, Washington DC}}},\ Vol.~\bibinfo {volume} {5}\ (\bibinfo
  {year} {2017})\ pp.\ \bibinfo {pages} {51--54}\BibitemShut {NoStop}%
\bibitem [{\citenamefont {Roy}(2015{\natexlab{a}})}]{roy14_5}%
  \BibitemOpen
  \bibfield  {author} {\bibinfo {author} {\bibfnamefont {K.}~\bibnamefont
  {Roy}},\ }\href@noop {} {\bibfield  {journal} {\bibinfo  {journal} {IEEE
  Trans. Magn.}\ }\textbf {\bibinfo {volume} {51}},\ \bibinfo {pages} {2500808}
  (\bibinfo {year} {2015}{\natexlab{a}})}\BibitemShut {NoStop}%
\bibitem [{\citenamefont {Sharrock}(1994)}]{RefWorks:2630}%
  \BibitemOpen
  \bibfield  {author} {\bibinfo {author} {\bibfnamefont {M.~P.}\ \bibnamefont
  {Sharrock}},\ }\href@noop {} {\bibfield  {journal} {\bibinfo  {journal} {J.
  Appl. Phys.}\ }\textbf {\bibinfo {volume} {76}},\ \bibinfo {pages} {6413}
  (\bibinfo {year} {1994})}\BibitemShut {NoStop}%
\bibitem [{\citenamefont {Weller}\ and\ \citenamefont
  {Moser}(1999)}]{RefWorks:2629}%
  \BibitemOpen
  \bibfield  {author} {\bibinfo {author} {\bibfnamefont {D.}~\bibnamefont
  {Weller}}\ and\ \bibinfo {author} {\bibfnamefont {A.}~\bibnamefont {Moser}},\
  }\href@noop {} {\bibfield  {journal} {\bibinfo  {journal} {IEEE Trans.
  Magn.}\ }\textbf {\bibinfo {volume} {35}},\ \bibinfo {pages} {4423} (\bibinfo
  {year} {1999})}\BibitemShut {NoStop}%
\bibitem [{\citenamefont {Richter}(2007)}]{RefWorks:1354}%
  \BibitemOpen
  \bibfield  {author} {\bibinfo {author} {\bibfnamefont {H.~J.}\ \bibnamefont
  {Richter}},\ }\href@noop {} {\bibfield  {journal} {\bibinfo  {journal} {J.
  Phys. D: Appl. Phys.}\ }\textbf {\bibinfo {volume} {40}},\ \bibinfo {pages}
  {R149} (\bibinfo {year} {2007})}\BibitemShut {NoStop}%
\bibitem [{\citenamefont {Sato}\ \emph {et~al.}(2014)\citenamefont {Sato},
  \citenamefont {Enobio}, \citenamefont {Yamanouchi}, \citenamefont {Ikeda},
  \citenamefont {Fukami}, \citenamefont {Kanai}, \citenamefont {Matsukura},\
  and\ \citenamefont {Ohno}}]{RefWorks:903}%
  \BibitemOpen
  \bibfield  {author} {\bibinfo {author} {\bibfnamefont {H.}~\bibnamefont
  {Sato}}, \bibinfo {author} {\bibfnamefont {E.~C.~I.}\ \bibnamefont {Enobio}},
  \bibinfo {author} {\bibfnamefont {M.}~\bibnamefont {Yamanouchi}}, \bibinfo
  {author} {\bibfnamefont {S.}~\bibnamefont {Ikeda}}, \bibinfo {author}
  {\bibfnamefont {S.}~\bibnamefont {Fukami}}, \bibinfo {author} {\bibfnamefont
  {S.}~\bibnamefont {Kanai}}, \bibinfo {author} {\bibfnamefont
  {F.}~\bibnamefont {Matsukura}}, \ and\ \bibinfo {author} {\bibfnamefont
  {H.}~\bibnamefont {Ohno}},\ }\href@noop {} {\bibfield  {journal} {\bibinfo
  {journal} {Appl. Phys. Lett.}\ }\textbf {\bibinfo {volume} {105}},\ \bibinfo
  {pages} {062403} (\bibinfo {year} {2014})}\BibitemShut {NoStop}%
\bibitem [{\citenamefont {Margulies}\ \emph {et~al.}(2004)\citenamefont
  {Margulies}, \citenamefont {Schabes}, \citenamefont {Supper}, \citenamefont
  {Do}, \citenamefont {Berger}, \citenamefont {Moser}, \citenamefont {Rice},
  \citenamefont {Arnett}, \citenamefont {Madison},\ and\ \citenamefont
  {Lengsfield}}]{RefWorks:1443}%
  \BibitemOpen
  \bibfield  {author} {\bibinfo {author} {\bibfnamefont {D.~T.}\ \bibnamefont
  {Margulies}}, \bibinfo {author} {\bibfnamefont {M.~E.}\ \bibnamefont
  {Schabes}}, \bibinfo {author} {\bibfnamefont {N.}~\bibnamefont {Supper}},
  \bibinfo {author} {\bibfnamefont {H.}~\bibnamefont {Do}}, \bibinfo {author}
  {\bibfnamefont {A.}~\bibnamefont {Berger}}, \bibinfo {author} {\bibfnamefont
  {A.}~\bibnamefont {Moser}}, \bibinfo {author} {\bibfnamefont {P.~M.}\
  \bibnamefont {Rice}}, \bibinfo {author} {\bibfnamefont {P.}~\bibnamefont
  {Arnett}}, \bibinfo {author} {\bibfnamefont {M.}~\bibnamefont {Madison}}, \
  and\ \bibinfo {author} {\bibfnamefont {B.}~\bibnamefont {Lengsfield}},\
  }\href@noop {} {\bibfield  {journal} {\bibinfo  {journal} {Appl. Phys.
  Lett.}\ }\textbf {\bibinfo {volume} {85}},\ \bibinfo {pages} {6200} (\bibinfo
  {year} {2004})}\BibitemShut {NoStop}%
\bibitem [{\citenamefont {Suess}\ \emph {et~al.}(2005)\citenamefont {Suess},
  \citenamefont {Schrefl}, \citenamefont {F\"{a}hler}, \citenamefont
  {Kirschner}, \citenamefont {Hrkac}, \citenamefont {Dorfbauer},\ and\
  \citenamefont {Fidler}}]{RefWorks:1449}%
  \BibitemOpen
  \bibfield  {author} {\bibinfo {author} {\bibfnamefont {D.}~\bibnamefont
  {Suess}}, \bibinfo {author} {\bibfnamefont {T.}~\bibnamefont {Schrefl}},
  \bibinfo {author} {\bibfnamefont {S.}~\bibnamefont {F\"{a}hler}}, \bibinfo
  {author} {\bibfnamefont {M.}~\bibnamefont {Kirschner}}, \bibinfo {author}
  {\bibfnamefont {G.}~\bibnamefont {Hrkac}}, \bibinfo {author} {\bibfnamefont
  {F.}~\bibnamefont {Dorfbauer}}, \ and\ \bibinfo {author} {\bibfnamefont
  {J.}~\bibnamefont {Fidler}},\ }\href@noop {} {\bibfield  {journal} {\bibinfo
  {journal} {Appl. Phys. Lett.}\ }\textbf {\bibinfo {volume} {87}},\ \bibinfo
  {pages} {012504} (\bibinfo {year} {2005})}\BibitemShut {NoStop}%
\bibitem [{\citenamefont {Suess}\ \emph {et~al.}(2007)\citenamefont {Suess},
  \citenamefont {Eder}, \citenamefont {Lee}, \citenamefont {Dittrich},
  \citenamefont {Fidler}, \citenamefont {Harrell}, \citenamefont {Schrefl},
  \citenamefont {Hrkac}, \citenamefont {Schabes},\ and\ \citenamefont
  {Supper}}]{RefWorks:1448}%
  \BibitemOpen
  \bibfield  {author} {\bibinfo {author} {\bibfnamefont {D.}~\bibnamefont
  {Suess}}, \bibinfo {author} {\bibfnamefont {S.}~\bibnamefont {Eder}},
  \bibinfo {author} {\bibfnamefont {J.}~\bibnamefont {Lee}}, \bibinfo {author}
  {\bibfnamefont {R.}~\bibnamefont {Dittrich}}, \bibinfo {author}
  {\bibfnamefont {J.}~\bibnamefont {Fidler}}, \bibinfo {author} {\bibfnamefont
  {J.~W.}\ \bibnamefont {Harrell}}, \bibinfo {author} {\bibfnamefont
  {T.}~\bibnamefont {Schrefl}}, \bibinfo {author} {\bibfnamefont
  {G.}~\bibnamefont {Hrkac}}, \bibinfo {author} {\bibfnamefont
  {M.}~\bibnamefont {Schabes}}, \ and\ \bibinfo {author} {\bibfnamefont
  {N.}~\bibnamefont {Supper}},\ }\href@noop {} {\bibfield  {journal} {\bibinfo
  {journal} {Phys. Rev. B}\ }\textbf {\bibinfo {volume} {75}},\ \bibinfo
  {pages} {174430} (\bibinfo {year} {2007})}\BibitemShut {NoStop}%
\bibitem [{\citenamefont {Richter}, \citenamefont {Girt},\ and\ \citenamefont
  {Zhou}(2002)}]{RefWorks:1447}%
  \BibitemOpen
  \bibfield  {author} {\bibinfo {author} {\bibfnamefont {H.~J.}\ \bibnamefont
  {Richter}}, \bibinfo {author} {\bibfnamefont {E.}~\bibnamefont {Girt}}, \
  and\ \bibinfo {author} {\bibfnamefont {H.}~\bibnamefont {Zhou}},\ }\href@noop
  {} {\bibfield  {journal} {\bibinfo  {journal} {Appl. Phys. Lett.}\ }\textbf
  {\bibinfo {volume} {80}},\ \bibinfo {pages} {2529} (\bibinfo {year}
  {2002})}\BibitemShut {NoStop}%
\bibitem [{\citenamefont {Richter}\ and\ \citenamefont
  {Dobin}(2006)}]{RefWorks:1446}%
  \BibitemOpen
  \bibfield  {author} {\bibinfo {author} {\bibfnamefont {H.~J.}\ \bibnamefont
  {Richter}}\ and\ \bibinfo {author} {\bibfnamefont {A.~Y.}\ \bibnamefont
  {Dobin}},\ }\href@noop {} {\bibfield  {journal} {\bibinfo  {journal} {J.
  Appl. Phys.}\ }\textbf {\bibinfo {volume} {99}},\ \bibinfo {pages} {08Q905}
  (\bibinfo {year} {2006})}\BibitemShut {NoStop}%
\bibitem [{\citenamefont {Choo}\ \emph {et~al.}(2007)\citenamefont {Choo},
  \citenamefont {Chantrell}, \citenamefont {Lamberton}, \citenamefont
  {Johnston},\ and\ \citenamefont {O'Grady}}]{RefWorks:1435}%
  \BibitemOpen
  \bibfield  {author} {\bibinfo {author} {\bibfnamefont {D.}~\bibnamefont
  {Choo}}, \bibinfo {author} {\bibfnamefont {R.~W.}\ \bibnamefont {Chantrell}},
  \bibinfo {author} {\bibfnamefont {R.}~\bibnamefont {Lamberton}}, \bibinfo
  {author} {\bibfnamefont {A.}~\bibnamefont {Johnston}}, \ and\ \bibinfo
  {author} {\bibfnamefont {K.}~\bibnamefont {O'Grady}},\ }\href@noop {}
  {\bibfield  {journal} {\bibinfo  {journal} {J. Appl. Phys.}\ }\textbf
  {\bibinfo {volume} {101}},\ \bibinfo {pages} {09E521} (\bibinfo {year}
  {2007})}\BibitemShut {NoStop}%
\bibitem [{\citenamefont {Bertram}\ and\ \citenamefont
  {Lengsfield}(2007)}]{RefWorks:1433}%
  \BibitemOpen
  \bibfield  {author} {\bibinfo {author} {\bibfnamefont {H.~N.}\ \bibnamefont
  {Bertram}}\ and\ \bibinfo {author} {\bibfnamefont {B.}~\bibnamefont
  {Lengsfield}},\ }\href@noop {} {\bibfield  {journal} {\bibinfo  {journal}
  {IEEE Trans. Mag.}\ }\textbf {\bibinfo {volume} {43}},\ \bibinfo {pages}
  {2145} (\bibinfo {year} {2007})}\BibitemShut {NoStop}%
\bibitem [{\citenamefont {Hauet}\ \emph {et~al.}(2009)\citenamefont {Hauet},
  \citenamefont {Dobisz}, \citenamefont {Florez}, \citenamefont {Park},
  \citenamefont {Lengsfield}, \citenamefont {Terris},\ and\ \citenamefont
  {Hellwig}}]{RefWorks:1441}%
  \BibitemOpen
  \bibfield  {author} {\bibinfo {author} {\bibfnamefont {T.}~\bibnamefont
  {Hauet}}, \bibinfo {author} {\bibfnamefont {E.}~\bibnamefont {Dobisz}},
  \bibinfo {author} {\bibfnamefont {S.}~\bibnamefont {Florez}}, \bibinfo
  {author} {\bibfnamefont {J.}~\bibnamefont {Park}}, \bibinfo {author}
  {\bibfnamefont {B.}~\bibnamefont {Lengsfield}}, \bibinfo {author}
  {\bibfnamefont {B.~D.}\ \bibnamefont {Terris}}, \ and\ \bibinfo {author}
  {\bibfnamefont {O.}~\bibnamefont {Hellwig}},\ }\href@noop {} {\bibfield
  {journal} {\bibinfo  {journal} {Appl. Phys. Lett.}\ }\textbf {\bibinfo
  {volume} {95}},\ \bibinfo {pages} {262504} (\bibinfo {year}
  {2009})}\BibitemShut {NoStop}%
\bibitem [{\citenamefont {Nolan}, \citenamefont {Valcu},\ and\ \citenamefont
  {Richter}(2011)}]{RefWorks:1444}%
  \BibitemOpen
  \bibfield  {author} {\bibinfo {author} {\bibfnamefont {T.~P.}\ \bibnamefont
  {Nolan}}, \bibinfo {author} {\bibfnamefont {B.~F.}\ \bibnamefont {Valcu}}, \
  and\ \bibinfo {author} {\bibfnamefont {H.~J.}\ \bibnamefont {Richter}},\
  }\href@noop {} {\bibfield  {journal} {\bibinfo  {journal} {IEEE Trans. Mag.}\
  }\textbf {\bibinfo {volume} {47}},\ \bibinfo {pages} {63} (\bibinfo {year}
  {2011})}\BibitemShut {NoStop}%
\bibitem [{\citenamefont {Nogu\'{e}s}\ \emph {et~al.}(2005)\citenamefont
  {Nogu\'{e}s}, \citenamefont {Sort}, \citenamefont {Langlais}, \citenamefont
  {Skumryev}, \citenamefont {{n}ach}, \citenamefont {{n}oz},\ and\
  \citenamefont {Bar\'{e}}}]{RefWorks:2587}%
  \BibitemOpen
  \bibfield  {author} {\bibinfo {author} {\bibfnamefont {J.}~\bibnamefont
  {Nogu\'{e}s}}, \bibinfo {author} {\bibfnamefont {J.}~\bibnamefont {Sort}},
  \bibinfo {author} {\bibfnamefont {V.}~\bibnamefont {Langlais}}, \bibinfo
  {author} {\bibfnamefont {V.}~\bibnamefont {Skumryev}}, \bibinfo {author}
  {\bibfnamefont {S.~S.}\ \bibnamefont {{n}ach}}, \bibinfo {author}
  {\bibfnamefont {J.~S.~M.}\ \bibnamefont {{n}oz}}, \ and\ \bibinfo {author}
  {\bibfnamefont {M.~D.}\ \bibnamefont {Bar\'{e}}},\ }\href@noop {} {\bibfield
  {journal} {\bibinfo  {journal} {Phys. Rep.}\ }\textbf {\bibinfo {volume}
  {422}},\ \bibinfo {pages} {65} (\bibinfo {year} {2005})}\BibitemShut
  {NoStop}%
\bibitem [{\citenamefont {Wang}, \citenamefont {Shen},\ and\ \citenamefont
  {Bai}(2005)}]{RefWorks:1456}%
  \BibitemOpen
  \bibfield  {author} {\bibinfo {author} {\bibfnamefont {J.~P.}\ \bibnamefont
  {Wang}}, \bibinfo {author} {\bibfnamefont {W.}~\bibnamefont {Shen}}, \ and\
  \bibinfo {author} {\bibfnamefont {J.}~\bibnamefont {Bai}},\ }\href@noop {}
  {\bibfield  {journal} {\bibinfo  {journal} {IEEE Trans. Mag.}\ }\textbf
  {\bibinfo {volume} {41}},\ \bibinfo {pages} {3181} (\bibinfo {year}
  {2005})}\BibitemShut {NoStop}%
\bibitem [{\citenamefont {Wang}, \citenamefont {Shen},\ and\ \citenamefont
  {Hong}(2007)}]{RefWorks:1457}%
  \BibitemOpen
  \bibfield  {author} {\bibinfo {author} {\bibfnamefont {J.~P.}\ \bibnamefont
  {Wang}}, \bibinfo {author} {\bibfnamefont {W.}~\bibnamefont {Shen}}, \ and\
  \bibinfo {author} {\bibfnamefont {S.~Y.}\ \bibnamefont {Hong}},\ }\href@noop
  {} {\bibfield  {journal} {\bibinfo  {journal} {IEEE Trans. Mag.}\ }\textbf
  {\bibinfo {volume} {43}},\ \bibinfo {pages} {682} (\bibinfo {year}
  {2007})}\BibitemShut {NoStop}%
\bibitem [{\citenamefont {Victora}\ and\ \citenamefont
  {Shen}(2005{\natexlab{a}})}]{RefWorks:1453}%
  \BibitemOpen
  \bibfield  {author} {\bibinfo {author} {\bibfnamefont {R.~H.}\ \bibnamefont
  {Victora}}\ and\ \bibinfo {author} {\bibfnamefont {X.}~\bibnamefont {Shen}},\
  }\href@noop {} {\bibfield  {journal} {\bibinfo  {journal} {IEEE Trans. Mag.}\
  }\textbf {\bibinfo {volume} {41}},\ \bibinfo {pages} {2828} (\bibinfo {year}
  {2005}{\natexlab{a}})}\BibitemShut {NoStop}%
\bibitem [{\citenamefont {Victora}\ and\ \citenamefont
  {Shen}(2005{\natexlab{b}})}]{RefWorks:1454}%
  \BibitemOpen
  \bibfield  {author} {\bibinfo {author} {\bibfnamefont {R.~H.}\ \bibnamefont
  {Victora}}\ and\ \bibinfo {author} {\bibfnamefont {X.}~\bibnamefont {Shen}},\
  }\href@noop {} {\bibfield  {journal} {\bibinfo  {journal} {IEEE Trans. Mag.}\
  }\textbf {\bibinfo {volume} {41}},\ \bibinfo {pages} {537} (\bibinfo {year}
  {2005}{\natexlab{b}})}\BibitemShut {NoStop}%
\bibitem [{\citenamefont {Victora}\ and\ \citenamefont
  {Shen}(2008)}]{RefWorks:1452}%
  \BibitemOpen
  \bibfield  {author} {\bibinfo {author} {\bibfnamefont {R.~H.}\ \bibnamefont
  {Victora}}\ and\ \bibinfo {author} {\bibfnamefont {X.}~\bibnamefont {Shen}},\
  }\href@noop {} {\bibfield  {journal} {\bibinfo  {journal} {Proc. IEEE}\
  }\textbf {\bibinfo {volume} {96}},\ \bibinfo {pages} {1799} (\bibinfo {year}
  {2008})}\BibitemShut {NoStop}%
\bibitem [{\citenamefont {Fullerton}\ \emph {et~al.}(2000)\citenamefont
  {Fullerton}, \citenamefont {Margulies}, \citenamefont {Schabes},
  \citenamefont {Carey}, \citenamefont {Gurney}, \citenamefont {Moser},
  \citenamefont {Best}, \citenamefont {Zeltzer}, \citenamefont {Rubin},\ and\
  \citenamefont {Rosen}}]{RefWorks:1439}%
  \BibitemOpen
  \bibfield  {author} {\bibinfo {author} {\bibfnamefont {E.~E.}\ \bibnamefont
  {Fullerton}}, \bibinfo {author} {\bibfnamefont {D.~T.}\ \bibnamefont
  {Margulies}}, \bibinfo {author} {\bibfnamefont {M.~E.}\ \bibnamefont
  {Schabes}}, \bibinfo {author} {\bibfnamefont {M.}~\bibnamefont {Carey}},
  \bibinfo {author} {\bibfnamefont {B.}~\bibnamefont {Gurney}}, \bibinfo
  {author} {\bibfnamefont {A.}~\bibnamefont {Moser}}, \bibinfo {author}
  {\bibfnamefont {M.}~\bibnamefont {Best}}, \bibinfo {author} {\bibfnamefont
  {G.}~\bibnamefont {Zeltzer}}, \bibinfo {author} {\bibfnamefont
  {K.}~\bibnamefont {Rubin}}, \ and\ \bibinfo {author} {\bibfnamefont
  {H.}~\bibnamefont {Rosen}},\ }\href@noop {} {\bibfield  {journal} {\bibinfo
  {journal} {Appl. Phys. Lett.}\ }\textbf {\bibinfo {volume} {77}},\ \bibinfo
  {pages} {3806} (\bibinfo {year} {2000})}\BibitemShut {NoStop}%
\bibitem [{\citenamefont {Tudosa}\ \emph {et~al.}(2010)\citenamefont {Tudosa},
  \citenamefont {Katine}, \citenamefont {Mangin},\ and\ \citenamefont
  {Fullerton}}]{RefWorks:1450}%
  \BibitemOpen
  \bibfield  {author} {\bibinfo {author} {\bibfnamefont {I.}~\bibnamefont
  {Tudosa}}, \bibinfo {author} {\bibfnamefont {J.~A.}\ \bibnamefont {Katine}},
  \bibinfo {author} {\bibfnamefont {S.}~\bibnamefont {Mangin}}, \ and\ \bibinfo
  {author} {\bibfnamefont {E.~E.}\ \bibnamefont {Fullerton}},\ }\href@noop {}
  {\bibfield  {journal} {\bibinfo  {journal} {Appl. Phys. Lett.}\ }\textbf
  {\bibinfo {volume} {96}},\ \bibinfo {pages} {212504} (\bibinfo {year}
  {2010})}\BibitemShut {NoStop}%
\bibitem [{\citenamefont {Yulaev}\ \emph {et~al.}(2011)\citenamefont {Yulaev},
  \citenamefont {Lubarda}, \citenamefont {Mangin}, \citenamefont {Lomakin},\
  and\ \citenamefont {Fullerton}}]{RefWorks:1125}%
  \BibitemOpen
  \bibfield  {author} {\bibinfo {author} {\bibfnamefont {I.}~\bibnamefont
  {Yulaev}}, \bibinfo {author} {\bibfnamefont {M.~V.}\ \bibnamefont {Lubarda}},
  \bibinfo {author} {\bibfnamefont {S.}~\bibnamefont {Mangin}}, \bibinfo
  {author} {\bibfnamefont {V.}~\bibnamefont {Lomakin}}, \ and\ \bibinfo
  {author} {\bibfnamefont {E.~E.}\ \bibnamefont {Fullerton}},\ }\href@noop {}
  {\bibfield  {journal} {\bibinfo  {journal} {Appl. Phys. Lett.}\ }\textbf
  {\bibinfo {volume} {99}},\ \bibinfo {pages} {132502} (\bibinfo {year}
  {2011})}\BibitemShut {NoStop}%
\bibitem [{\citenamefont {Cuchet}\ \emph {et~al.}(2015)\citenamefont {Cuchet},
  \citenamefont {Rodmacq}, \citenamefont {Auffret}, \citenamefont {Sousa},
  \citenamefont {Prejbeanu},\ and\ \citenamefont {Di\'{e}ny}}]{RefWorks:1119}%
  \BibitemOpen
  \bibfield  {author} {\bibinfo {author} {\bibfnamefont {L.}~\bibnamefont
  {Cuchet}}, \bibinfo {author} {\bibfnamefont {B.}~\bibnamefont {Rodmacq}},
  \bibinfo {author} {\bibfnamefont {S.}~\bibnamefont {Auffret}}, \bibinfo
  {author} {\bibfnamefont {R.~C.}\ \bibnamefont {Sousa}}, \bibinfo {author}
  {\bibfnamefont {I.~L.}\ \bibnamefont {Prejbeanu}}, \ and\ \bibinfo {author}
  {\bibfnamefont {B.}~\bibnamefont {Di\'{e}ny}},\ }\href@noop {} {\bibfield
  {journal} {\bibinfo  {journal} {J. Appl. Phys.}\ }\textbf {\bibinfo {volume}
  {117}},\ \bibinfo {pages} {233901} (\bibinfo {year} {2015})}\BibitemShut
  {NoStop}%
\bibitem [{\citenamefont {Cl\'{e}ment}\ \emph {et~al.}(2015)\citenamefont
  {Cl\'{e}ment}, \citenamefont {Baraduc}, \citenamefont {Ducruet},
  \citenamefont {Vila}, \citenamefont {Chshiev},\ and\ \citenamefont
  {Di\'{e}ny}}]{RefWorks:1120}%
  \BibitemOpen
  \bibfield  {author} {\bibinfo {author} {\bibfnamefont {P.~Y.}\ \bibnamefont
  {Cl\'{e}ment}}, \bibinfo {author} {\bibfnamefont {C.}~\bibnamefont
  {Baraduc}}, \bibinfo {author} {\bibfnamefont {C.}~\bibnamefont {Ducruet}},
  \bibinfo {author} {\bibfnamefont {L.}~\bibnamefont {Vila}}, \bibinfo {author}
  {\bibfnamefont {M.}~\bibnamefont {Chshiev}}, \ and\ \bibinfo {author}
  {\bibfnamefont {B.}~\bibnamefont {Di\'{e}ny}},\ }\href@noop {} {\bibfield
  {journal} {\bibinfo  {journal} {Appl. Phys. Lett.}\ }\textbf {\bibinfo
  {volume} {107}},\ \bibinfo {pages} {102405} (\bibinfo {year}
  {2015})}\BibitemShut {NoStop}%
\bibitem [{\citenamefont {Roy}(2015{\natexlab{b}})}]{roy15_1}%
  \BibitemOpen
  \bibfield  {author} {\bibinfo {author} {\bibfnamefont {K.}~\bibnamefont
  {Roy}},\ }\href@noop {} {\bibfield  {journal} {\bibinfo  {journal} {Sci.
  Rep.}\ }\textbf {\bibinfo {volume} {5}},\ \bibinfo {pages} {10822} (\bibinfo
  {year} {2015}{\natexlab{b}})}\BibitemShut {NoStop}%
\bibitem [{\citenamefont {Landau}\ and\ \citenamefont
  {Lifshitz}(1935)}]{RefWorks:162}%
  \BibitemOpen
  \bibfield  {author} {\bibinfo {author} {\bibfnamefont {L.}~\bibnamefont
  {Landau}}\ and\ \bibinfo {author} {\bibfnamefont {E.}~\bibnamefont
  {Lifshitz}},\ }\href@noop {} {\bibfield  {journal} {\bibinfo  {journal}
  {Phys. Z. Sowjet.}\ }\textbf {\bibinfo {volume} {8}},\ \bibinfo {pages} {153}
  (\bibinfo {year} {1935})}\BibitemShut {NoStop}%
\bibitem [{\citenamefont {Gilbert}(2004)}]{RefWorks:161}%
  \BibitemOpen
  \bibfield  {author} {\bibinfo {author} {\bibfnamefont {T.~L.}\ \bibnamefont
  {Gilbert}},\ }\href@noop {} {\bibfield  {journal} {\bibinfo  {journal} {IEEE
  Trans. Magn.}\ }\textbf {\bibinfo {volume} {40}},\ \bibinfo {pages} {3443}
  (\bibinfo {year} {2004})}\BibitemShut {NoStop}%
\bibitem [{\citenamefont {Brown}(1963)}]{RefWorks:186}%
  \BibitemOpen
  \bibfield  {author} {\bibinfo {author} {\bibfnamefont {W.~F.}\ \bibnamefont
  {Brown}},\ }\href@noop {} {\bibfield  {journal} {\bibinfo  {journal} {Phys.
  Rev.}\ }\textbf {\bibinfo {volume} {130}},\ \bibinfo {pages} {1677} (\bibinfo
  {year} {1963})}\BibitemShut {NoStop}%
\bibitem [{\citenamefont {Parkin}, \citenamefont {More},\ and\ \citenamefont
  {Roche}(1990)}]{RefWorks:432}%
  \BibitemOpen
  \bibfield  {author} {\bibinfo {author} {\bibfnamefont {S.~S.~P.}\
  \bibnamefont {Parkin}}, \bibinfo {author} {\bibfnamefont {N.}~\bibnamefont
  {More}}, \ and\ \bibinfo {author} {\bibfnamefont {K.~P.}\ \bibnamefont
  {Roche}},\ }\href@noop {} {\bibfield  {journal} {\bibinfo  {journal} {Phys.
  Rev. Lett.}\ }\textbf {\bibinfo {volume} {64}},\ \bibinfo {pages} {2304}
  (\bibinfo {year} {1990})}\BibitemShut {NoStop}%
\bibitem [{\citenamefont {Parkin}(1991)}]{RefWorks:2632}%
  \BibitemOpen
  \bibfield  {author} {\bibinfo {author} {\bibfnamefont {S.~S.~P.}\
  \bibnamefont {Parkin}},\ }\href@noop {} {\bibfield  {journal} {\bibinfo
  {journal} {Phys. Rev. Lett.}\ }\textbf {\bibinfo {volume} {67}},\ \bibinfo
  {pages} {3598} (\bibinfo {year} {1991})}\BibitemShut {NoStop}%
\end{thebibliography}
\end{document}